\documentclass[preprint2]{emulateapj}
\shorttitle{Strong recombination structures in W49B}
\shortauthors{Ozawa et al.}

\begin{document}
\title{{\it Suzaku} Discovery of the Strong Radiative Recombination Continuum of Iron\\ 
from the Supernova Remnant W49B}

\author{M. Ozawa\altaffilmark{1}}
\email{midori@cr.scphys.kyoto-u.ac.jp}
\author{K. Koyama\altaffilmark{1}}
\author{H. Yamaguchi\altaffilmark{2}}
\author{K. Masai\altaffilmark{3}}
\author{T. Tamagawa\altaffilmark{2}}
\altaffiltext{1}{Department of Physics, Kyoto University, Kitashirakawa-oiwake-cho, Sakyo-ku, Kyoto 606-8502, Japan}
\altaffiltext{2}{RIKEN (The Institute of Physical and Chemical Research),  2-1 Hirosawa, Wako, Saitama 351-0198, Japan}
\altaffiltext{3}{Department of Physics, Tokyo Metropolitan University, 1-1 Minami-Osawa, Hachioji, Tokyo 192-0397, Japan}

\begin{abstract}
We present a  hard X-ray spectrum of unprecedented quality 
of the Galactic supernova remnant W49B obtained with the {\it Suzaku} satellite. 
The spectrum exhibits an unusual structure consisting of a saw-edged bump above 8~keV. 
This bump cannot be explained by any combination of high-temperature plasmas in ionization equilibrium.
We firmly conclude that this bump is caused by the strong radiative recombination 
continuum (RRC) of iron, detected for the first time in a supernova remnant. 
The electron temperature derived from the bremsstrahlung continuum shape 
and the slope of the RRC is $\sim$1.5~keV. 
On the other hand, the ionization temperature derived from the observed intensity 
ratios between the RRC and K$\alpha$ lines of iron is $\sim$2.7~keV. 
These results indicate that the plasma is in a highly overionized state.
Volume emission measures independently determined from the fluxes of the thermal 
and RRC components are consistent with each other, 
suggesting the same origin of these components.
\end{abstract}
\keywords{ISM: individual (W49B) --- supernova remnants 
--- radiation mechanisms: thermal --- X-rays: ISM}

\section{Introduction} \label{sec:introduction}
W49B (G43.3-0.2) is a Galactic supernova remnant (SNR) with 
strong X-ray line emissions from highly ionized atoms. 
It exhibits centrally filled X-rays inside a bright radio shell with a radius of 100 arcsec (Pye et al. 1984).
The distance to W49B remains uncertain.
It was estimated to lie at a distance of $\sim$8~kpc (Radhakrishnan et al. 1972; Moffett \& Reynolds 1994).
Using the spectral distribution of HI absorption, however, Brogan \& Troland (2001) found no clear evidence 
that W49B is closer to the sun than W49A,
which is located at a distance of $\sim$11.4~kpc (Gwinn et al. 1992). 
This may indicate a possible association of W49B with the star-forming region W49A.
The near-infrared narrowband images indicate a barrel-shaped structure with coaxial rings, 
which is suggestive of bipolar wind structures surrounding massive stars (Keohane et al. 2007). 
They also showed an X-ray image from {\it Chandra}, which has a jet-like structure along the axis of the barrel.
They interpreted these findings as evidence that W49B had exploded inside a wind-blown 
bubble in a dense molecular cloud.

Using {\it ASCA} data, Hwang et al. (2000) showed 
that broadband modeling of the remnant's spectrum 
required two thermal components (0.2~keV and 2~keV) 
and significant overabundances of Si, S, Ar, Ca, Fe, and Ni.
They confirmed that most of the X-ray emitting plasma was nearly in collisional ionization equilibrium (CIE). 
They also found evidence for Cr and Mn K$\alpha$ emission.

Kawasaki et al.\ (2005) claimed the presence of ``overionized" plasma in W49B 
through the analysis of {\it ASCA} 2.75--6.0~keV spectrum.
They measured intensity ratios of the H-like K$\alpha$ (hereafter Ly$\alpha$) 
to He-like K$\alpha$ (He$\alpha$) lines of Ar and Ca to obtain the ionization 
temperature ($kT_z$), and found that $kT_z$ ($\sim$2.5~keV) is significantly 
higher than the electron temperature ($kT_{\rm e}$) determined from the 
bremsstrahlung continuum shape ($\sim$1.8~keV). 
Miceli et al.\ (2006) adopted the same analysis procedure for the {\it XMM-Newton} 
spectrum of the central region but found no evidence for the overionized state.

In this letter, we report the firm evidence for overionized plasma in W49B
using data from the X-ray Imaging Spectrometers (XIS: Koyama et al.\  2007)  
onboard the {\it Suzaku} satellite (Mitsuda et al.\ 2007).

\begin{figure}[t]
\includegraphics[scale=.60]{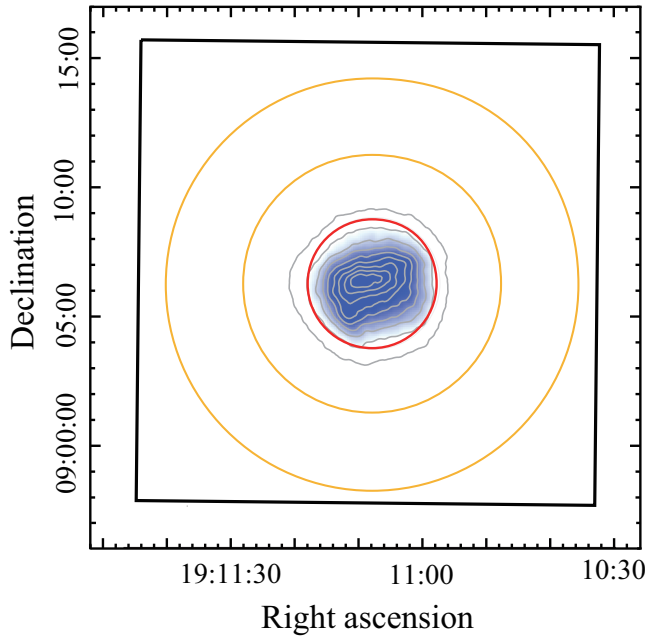}
\caption{Vignetting-corrected XIS image of W49B in the 1.5--7~keV band shown 
on a linear intensity scale. Data from the three active XISs are combined. 
Gray contours indicate every 10\% intensity level relative to the peak surface brightness. 
The red circle and orange annulus indicate the source and background regions, respectively. 
The XIS field of view is shown by the black square. 
  \label{fig:image}}
\end{figure}

\section{Observation and Data Reduction} \label{sec:observation}
We observed W49B with {\it Suzaku} on 2009 March 29 and 31 
(observation IDs 503084010 and 504035010, respectively). 
Because the pointing positions and rotation angles do not differ significantly, we merged these two data sets. 
The XIS consists of four X-ray CCD camera systems placed on the 
focal planes of four X-ray telescopes (Serlemitsos et al.\ 2007).
Three of them were operated
in the normal full-frame clocking mode with a spaced-row charge injection technique (Uchiyama et al. 2009)
during our observations. 
Two are front-illuminated (FI) CCDs and the other is a back-illuminated (BI) CCD.
Data were cleaned using processing version 2.2.11.24.
We used HEASOFT version 6.5.1 for data reduction and 
XSPEC version 11.3.2 for spectral analysis.
After screening with the standard criteria,
\footnote{http://heasarc.nasa.gov/docs/suzaku/processing/criteria\_xis.html}
the net integration time was $\sim$113~ksec. 

\section{Analysis} \label{sec:result}
\subsection{Construction of the Spectrum}
\begin{figure}[t]
\includegraphics[scale=.38]{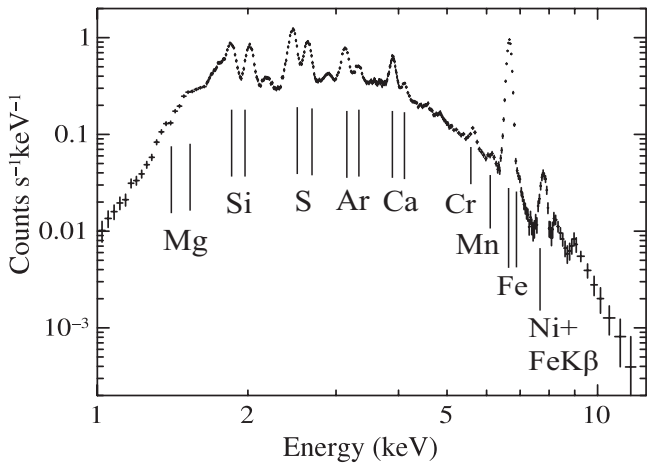}
\caption{Background-subtracted XIS FI spectrum. 
The energies of the prominent emission lines from specific elements 
are labeled. 
\label{fig:full}}
\end{figure}

Figure~\ref{fig:image} shows the vignetting-corrected XIS image in 
1.5--7~keV, the energy band including the major lines of K-shell 
emissions from Si, S, Ar, Ca, and Fe. 
We extracted the source spectrum from a circular region
with a radius of 2.5~arcmin and the background spectrum from an annulus 
surrounding the source with inner and outer radii of 5 and 8~arcmin, respectively. 
Non X-ray background (NXB) spectra constructed with the \texttt{xisnxbgen} software
were subtracted from both the source and background data. 
After correcting the vignetting and accumulation area, the source spectrum 
was made by subtracting the background spectrum. The spectra of the two FI 
sensors were merged to improve the statistics because their response functions 
were almost identical. 
Figure~\ref{fig:full} shows the merged FI spectrum. We can see several 
prominent lines of K-shell emission from He- and H-like ions.

\subsection{APEC Model Fit in the Hard X-ray Band}
As shown in Figure~\ref{fig:full}, the wide-band spectrum includes too 
many emission lines and possibly other complicated structures.  
For clean analyses and discussion, we focus on 
the 5--12~keV energy band, where emission lines and other structures mainly originate from Fe. 
We used only the FI data because
the NXB count rate of the BI data is much larger than the source rate in this energy band.

We first fitted the spectrum with a VAPEC (CIE plasma) model (Smith et al.\ 2001). 
The Fe and Ni abundances (hereafter $Z_{\rm Fe}$ and $Z_{\rm Ni}$)
normalized by the number fraction of the solar photosphere 
(Anders \& Grevesse 1989)  were free parameters. 
Since the Cr and Mn K$\alpha$ emission lines are not included in the VAPEC 
model, we added these lines with Gaussian functions. 
We fixed the interstellar extinction to a hydrogen column density of 
$5 \times 10^{22}$~cm$^{-2}$ with the solar elemental abundances, 
following Hwang et al. (2000).   
To fine tune the energy scale, an offset was added as a free parameter 
and found to be --4.7~eV, which was within the allowable range of the 
calibration uncertainties. 
We added this offset for further spectral fitting. 
Figure~\ref{fig:fit}a shows the fitting result of the one-VAPEC model, 
with a temperature of 1.64~keV. 
The model exhibited significant excess around the Fe-Ly$\alpha$ line and above 8~keV, 
and hence was rejected with a large $\chi^2$/dof (degree of freedom) of 1051/138. 

We then tried a two-VAPEC model, assuming equal abundances between the two components. 
Fe-Ly$\alpha$ was successfully reproduced, 
but the temperature of one component was unreasonably high ($\sim$70~keV), 
and a large residual above 8~keV remained with an unacceptable $\chi^2$/dof of 562/136.
Adding a further plasma or power-law component did not improve the fitting 
any more. In every case, the large bump above 8~keV was present. 

To examine whether this bump is real or artificial, 
we checked the light curves of the source and background regions in the 8.5--10~keV band. 
They showed almost constant fluxes,
indicating no flare-like event had occurred during the observations.  
In addition, we detected the bump in both FI spectra (XIS~0 and XIS~3) 
and even in the BI spectrum. 
Thus, the bump is a real structure. 

The edge energy of the bump ($\sim$9~keV) corresponds to the electron binding energy of Fe. 
This suggests that the saw-edged bump is likely due to 
a radiative recombination continuum (RRC): X-ray emissions due to the 
free-bound transition of electrons.
Hereafter, we call the RRC accompanied by the recombination of H-like ions 
into the ground state of He-like ions the He-RRC, 
and that of fully ionized ions into the ground state of H-like ions the H-RRC. 

\subsection{Recombination Continuum and Lines}
Both the line-like excess around Fe-Ly$\alpha$ and the bump above 8~keV are suggestive of 
the overionized state
because there should be a greater fraction of H-like ions in the overionized state than in the CIE.
This excessive number of H-like ions causes strong He-RRC and Fe-Ly$\alpha$ emissions.
However, no current plasma code can be applied to the overionized plasma. 
We, therefore, introduce the Fe-Ly$\alpha$ line and recombination structures 
consisting of the RRC and several emission lines below the K-edge energy of the RRC ($E_{\rm edge}$),
in addition to the one-VAPEC model.
We consider both He-RRC and H-RRC for a consistency check (see Section~4.2). 

We assume that the RRC is expressed as 
\begin{equation}
\frac{dN}{dE} \propto {\rm exp}\left( -\frac{E - E_{\rm edge}}{kT_{\rm e}}\right), 
~~~ {\rm for}~ E \geq E_{\rm edge~}~, 
\label{eq:edge}
\end{equation}
where $kT_{\rm e}$ is the electron temperature of the relevant recombining plasma. 
This formula gives a good approximation when the electron temperature is 
much lower than the K-edge energy ($kT_{\rm e} \ll E_{\rm edge}$) 
(e.g., Smith \& Brickhouse 2002).
In the case of W49B, $kT_{\rm e}$ is $\sim$1.5~keV, as we confirm later, 
while $E_{\rm edge}$ of Fe is $\sim$9~keV. 
Thus, we can safely adopt this formula. 

The recombination lines originate from cascade decays of electrons
that are captured into the excited levels of ions by the free-bound transition. 
We designate the cascade lines of He-like Fe  
as He$\alpha_{\rm rec}$ (the principle quantum number $n$ = 2$\to$1) and He$\beta_{\rm rec}$ (3$\to$1). 
For other lines from higher levels ($n \ge$ 4$\to$1), 
we combine them into one broad Gaussian function (He$\gamma$-$\infty_{\rm rec}$). 
The line widths and center energies of He$\alpha_{\rm rec}$ and 
He$\beta_{\rm rec}$ are fixed at zero and the experimental 
values for a charge exchange process, respectively (Wargelin et al.\ 2005). 
On the other hand, those of He$\gamma$-$\infty_{\rm rec}$ are allowed to vary freely. 
The fluxes of all these lines are free parameters.  

\begin{figure}[t]
\includegraphics[scale=.60]{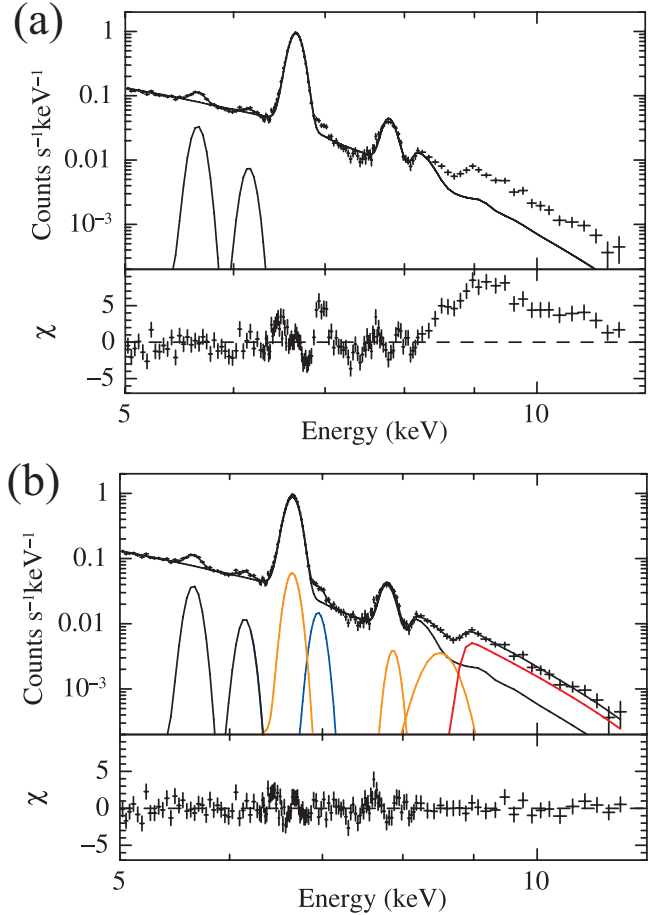}
\caption{(a) XIS spectrum in the 5--12~keV band. 
  The best-fit VAPEC model and additional K$\alpha$ lines 
  of Cr and Mn with Gaussian functions are shown by solid lines.
  The lower panel shows the residual from the best-fit model. \ 
  (b) Same spectrum as (a), but with the radiative recombination continuum of 
  He-like Fe (red), recombination lines (orange), and a Ly$\alpha$ line (blue). \label{fig:fit}}
\end{figure}

The resultant best-fit parameters and models are given in Table~\ref{tab:fit} and 
Figure~\ref{fig:fit}b, respectively. 
With this model, $\chi ^2$/dof is greatly improved to 193/128, 
although this is still unacceptable in a purely statistical sense. 
We can see significant data excesses at the lower energy sides of the 
He$\alpha$ lines of Fe ($\sim$ 6.5~keV) and Ni ($\sim$ 7.6~keV).
The former is likely due to the incomplete response function. 
We should note that Fe-He$\alpha$ statics are superior to
those for any other objects observed with {\it Suzaku}, and no significant residual has appeared so far.
If we observe carefully, we find a similar feature in the Mn-K$\alpha$ line from the onboard calibration source.
The latter would be partially due to the same reason given above, 
but is mainly due to lack of satellite lines of the Be-like and lower ionization 
states\footnote{As for Fe, such satellite lines are included in the VAPEC model.} of Ni in the VAPEC model. 
If we ignore the 6.4--6.6~keV and 7.6--7.7~keV band to escape these effects, 
an acceptable $\chi ^2$/dof of 118/107 is obtained.

\begin{table*}[t]
\begin{center}
\caption{Best-fit spectral parameters  \label{tab:fit}}
\begin{tabular}{lllll}
  \tableline\tableline
 CIE (VAPEC)& $kT_{\rm e}$ [keV]   & $Z_{\rm Fe}$  [solar] & $Z_{\rm Ni}$  [solar]& Normalization\tablenotemark{a}
\\  \cline{2-5}
 & 1.52 (1.50--1.53)  & 4.44 (4.36--4.53)\tablenotemark{b}& 10.9 (9.14--12.7)\tablenotemark{b} & 1.61 (1.60--1.62)\\
\hline\hline
 Line  &  & Center energy [keV]& Line width [keV] & Normalization\tablenotemark{c}  \\  \cline{2-5}
         & Cr K$\alpha$  &  5.655 (5.646--5.663) & 0 (fixed) &  2.85 ~(2.59--3.11)\\
     ~ & Mn K$\alpha$  & 6.162 (6.142--6.183) & 0 (fixed) &   0.957  (0.750--1.16) \\
  ~ & Fe He$\alpha _{\rm rec}$\tablenotemark{d}  & 6.666 (fixed\tablenotemark{e})  & 0 (fixed) &     5.30 ~(4.20--6.39)\\ 
      ~ & Fe Ly$\alpha$  & 6.961 (6.946--6.971) & 0 (fixed) &      1.46 ~(1.27--1.65) \\
   ~ & Fe He$\beta_{\rm rec}$\tablenotemark{d}  & 7.880 (fixed\tablenotemark{e})  & 0 (fixed) &     0.593 (0.338--0.847) \\
~ & Fe He$\gamma$-$\infty_{\rm rec}$\tablenotemark{d}& 8.538 (8.492--8.585) 
                                                                                   &0.200 (0.163--0.246) & 2.12 ~(1.81--2.43) \\ \hline\hline
RRC &    & Edge energy [keV]& $kT_{\rm e}$ [keV] & Normalization\tablenotemark{c}  \\  \cline{2-5}
& Fe He-RRC& 8.830 (fixed) &  1.43 (1.30--1.59)\tablenotemark{f}  & 11.2 ~(10.4--11.9)    \\
& Fe H-RRC& 9.194 (fixed)& 1.43 (1.30--1.59)\tablenotemark{f}     & 0.257 (0.00--1.04)\\  
\tableline\tableline
\end{tabular}
\end{center}
\tablecomments{
 The uncertainties in the parentheses are the 90\% confidence range. 
  $^{a}$The unit is $10^{-13}\int n_e n_{\rm H}~dV$/(4$\pi D^2$)  (cm$^{-5}$), 
  where $n_e$, $n_{\rm H}$, $V$, and $D$ are the electron and hydrogen densities (cm$^{-3}$), 
 emitting volume (cm$^3$), and distance to the source (cm), respectively. 
  $^{b}$These values should be modified in the case of the overionized plasma. See Section 4.3 for details.   
  $^{c}$The unit is $10^{-5}$ cm$^{-2}$ s$^{-1}$.   
  $^{d}$Lines emitted by a recombination process. 
   $^{e}$Fixed at the experimental values (Wargelin et al.\ 2005).
   $^{f}$We assumed the same $kT_{\rm e}$.}
\end{table*}

\section{Results and Discussion} \label{sec:discussion}
We have found, for the first time, the strong He-RRC of Fe from W49B.
Yamaguchi et al.\ (2009) recently discovered the H-RRC of Si and S from a middle-aged SNR, IC~443, 
and hence this is the second discovery of a clear RRC from an SNR. 
We also discovered RRC-accompanied recombination lines,
which may provide good diagnostics for the overionized plasma.
We review a quantitative verification of our spectral 
analysis and discuss the implications of the results. 

\subsection{Contribution of the Recombination Lines}
We discuss the validity of the best-fit fluxes of the RRC and recombination lines in Table~1. 
The recombination cross section of the H-like ions into a level of $n$ 
is approximately given as shown below (e.g., Nakayama \& Masai \ 2001).
\begin{equation}
\sigma_{n} \propto \frac{1}{n^3} \bigg(\frac{3}{2} \frac{kT_{\rm e}}{E_{\rm edge}}+ \frac{1}{n^2} \bigg) ^{-1}
\label{eq:RC-line}
\end{equation}
We apply this approximation for the He-like ions, but $\sigma_{1}$ must be reduced by half  
because one electron is already at the ground state. 
In principle, the recombination line flux can be estimated by the branching ratio to various levels, 
but these processes are very complicated. We, therefore, base our discussion only on a simple picture.

We compare the predicted capture and  observed transition rates normalized with the $n$ = 1 value.
We can estimate the capture rates using Equation~2 as $\sigma_2/\sigma_1$ = 0.62, 
$\sigma_3/\sigma_1$ = 0.25, and $(\sigma_4+\sigma_5+...)/\sigma_1$ = 0.34.
On the other hand, the observed flux rates are
He$\alpha_{\rm rec}$/He-RRC = 0.47 ($\pm$0.10), He$\beta_{\rm rec}$/He-RRC = 0.053 ($\pm$0.023), 
and He$\gamma$-$\infty_{\rm rec}$/He-RRC = 0.19 
($\pm$0.03).\footnote{Throughout this paper, all the errors in the parentheses are at 90\% confidence level.}
By taking the fractions between the observed and predicted rates,
we obtain 76 ($\pm$16)\%, 21 ($\pm$9)\%, and 56 ($\pm$9)\% 
for the He$\alpha_{\rm rec}$, He$\beta_{\rm rec}$, and He$\gamma$-$\infty_{\rm rec}$ lines, respectively.

These fractions for the He$\beta_{\rm rec}$ and He$\gamma$-$\infty_{\rm rec}$ lines may be conceivable, 
because one electron at $n$ = 1 suppresses
direct transitions from excited levels ($n \ge$ 3$\to$1) for the He-like ions. 
The fraction of He$\alpha_{\rm rec}$ is slightly smaller than expected because 
there should be a significant contribution of the cascade decay electrons from higher levels ($n \ge$ 3$\to$2$\to$1).
The real He$\alpha_{\rm rec}$ flux may be somewhat larger due to the uncertainty of the response function,
but the contribution of the He$\alpha_{\rm rec}$ flux relative to the total He$\alpha$ is  only 
$\lesssim$10\% and does not affect results and following discussion.    

\subsection{Electron and Ionization Temperatures}
The electron temperatures determined by the bremsstrahlung continuum shape and the RRC slope
are 1.52 (1.50--1.53) keV and 1.43 (1.30--1.59) keV, respectively. 
These consistent results indicate a common origin of these emissions.

The ionization temperatures directly reflect the ion fractions 
and hence can be determined as shown below.
From the best-fit model in Table~1, the flux ratios of
Ly$\alpha$/He$\alpha$, He-RRC/He$\alpha$, and H-RRC/He-RRC are given as
0.016 (0.014--0.018), 0.12 (0.11--0.13), and 0.023 ($\le$0.10).
In Figure~4, we compare these values with the modeled emissivity ratios 
derived from the radiation code of Masai (1994) for a plasma of $kT_{\rm e}$ = 1.5~keV.
We obtain $kT_z$ = 2.58 (2.46--2.68)~keV, 2.68 (2.63--2.73) keV, 
and 2.55 ($\le$3.65) keV, respectively, for the above ratios. 
We also compare the Ly$\alpha$/He$\alpha$ ratio 
with that derived from the APEC code (Smith et al. 2001), 
although this code is valid only for a CIE state.
We obtain $kT_z$ = 2.46 (2.39--2.54) keV, which is within the margin of error of the above results.  
The ionization temperatures ($\sim$2.7~keV) are significantly higher 
than the electron temperatures ($\sim$1.5~keV),
indicating that the plasma is in a highly overionized state.

The first hint of overionized plasma in W49B was found by Kawasaki et al. (2005). 
Although they analyzed different elements (Ar and Ca) in different energy bands (2.75--6.0~keV),
and derived $kT_z$ from the Ly$\alpha$/He$\alpha$ ratio using the CIE plasma code, 
their results ($kT_z$ $\sim$2.5~keV and $kT_{\rm e}$ $\sim$1.8~keV) are nearly consistent with ours.
Our claim is more essential because it is based on clear detection of the recombination structures.

\subsection{Iron and Nickel Abundances}
The abundances listed in Table~1 are valid only for the CIE state
and should be modified in the overionized case.
Since no plasma code can be applied to the overionized plasma currently,
we make possible modifications using an available APEC code. 

The He$\alpha$ intensity is proportional to $Z\times \epsilon(kT_{\rm e}, kT_z)$,
where $Z$ and $\epsilon(kT_{\rm e}, kT_z)$ are the abundance of the element (solar) 
and the total emissivity for the He$\alpha$ for $kT_{\rm e}$ and $kT_z$ (cm$^3$s$^{-1}$), respectively.
The emissivities of the He-, Li-, Be-, and B-like ions for Fe and the He- and Li-like ions for Ni
are modified by multiplying the ion-fraction ratio 
between $kT_z$ = 2.7~keV and 1.5~keV (Mazzotta et al. 1998). 
The total emissivity is given by adding those in individual ionization states.
Multiplying by $\epsilon$(1.5~keV, 1.5~keV)/$\epsilon$(1.5~keV, 2.7~keV), 
we obtain the real abundances in the overionized state as 
$Z_{\rm Fe}$$\sim$4.9~solar and $Z_{\rm Ni}$$\sim$5.2~solar.
Both the elements are highly over abundant, indicating an ejecta origin of the plasma. 

\subsection{Volume Emission Measure}
To check the consistency of the common origin of the bremsstrahlung and RRC emissions, 
we compare the volume emission measure (VEM).
The VEM is given by  $\int n_e n_{\rm H} dV/(4\pi D^2)$, 
where $n_e$, $n_{\rm H}$, $V$, 
and $D$ are the electron and hydrogen densities (cm$^{-3}$),
emitting volume (cm$^3$), and distance to the source (cm), respectively.  

The VEM of the VAPEC component ($VEM_{\rm VAPEC}$) is derived from Table~1
as  $VEM_{\rm VAPEC}$ = 1.61 (1.60 -- 1.62) $\times 10^{13}$ cm$^{-5}$.
On the other hand, the VEM of the RRC plasma ($VEM_{\rm RRC}$) is calculated from
\begin{equation}
F_{\rm RRC}=\alpha _1(kT_{\rm e}) \times \frac{n_{\rm Fe}}{n_{\rm H}} \times 
\kappa_{\rm H-like}(kT_z) \times VEM_{\rm RRC},
\label{eq:rrcflux}
\end{equation}
where $F_{\rm RRC}$, $\alpha _1(kT_{\rm e})$, $n_{\rm Fe}$, and $\kappa_{\rm H-like}(kT_z)$ are 
the flux of the He-RRC (cm$^{-2}$s$^{-1}$), 
the K-shell recombination rate coefficient for $kT_{\rm e}$ (cm$^{3}$s$^{-1}$), 
the number density of Fe (cm$^{-3}$), 
and the ion fraction of H-like Fe for $kT_z$, respectively. 
According to Badnell (2006), 
the total radiative recombination rate of He-like Fe at $kT_{\rm e}$ = 1.5~keV is given as 
$\sim$3.9$\times 10^{-12}$ cm$^{3}$s$^{-1}$.
The recombination rate into the ground state is given using Equation~2
as $\sigma_{1}/(\sigma_1 + \sigma_2 +...)\sim$0.45.
The value of $n_{\rm Fe}/n_{\rm H}$ is calculated using $Z_{\rm Fe}$ in Section 4.3 
and the number density of the solar photosphere (Anders \& Grevesse 1989).
Using the observed He-RRC flux,
we obtain\footnote{Here, we consider that the $kT_z$ error is 2.4--2.8~keV.
The large error of $VEM_{\rm RRC}$ is due to this effect.}
 $VEM_{\rm RRC}$ = 0.81 (0.58 -- 1.60)$\times 10^{13}$ cm$^{-5}$.
The two independent estimations of VEM give consistent results, supporting the same origin of the 
overionized plasma in W49B.

The origin of the overionized plasma in SNRs is an open question, and beyond the scope of this paper.
We simply note the possibility of the cooling of electrons via thermal conduction (Kawasaki et al. 2005) or
a more drastic cooling caused when the blast wave breaks out of some ambient matter 
into the rarefied interstellar medium (Yamaguchi et al. 2009).
If the latter is the case, a massive progenitor that had blown a stellar wind is likely to be favored.

\begin{figure}[t]
\includegraphics[scale=.60]{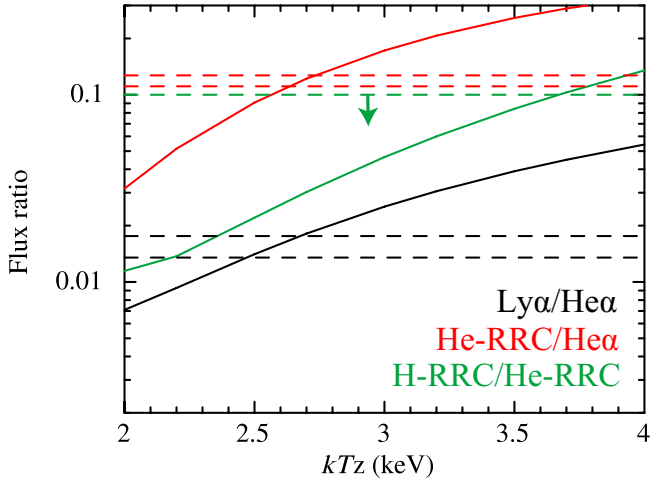}
\caption{Predicted emissivity ratios of Ly$\alpha$/He$\alpha$ (black), He-RRC/He$\alpha$ (red), 
and H-RRC/He-RRC (green) of Fe as a function of the ionization temperature ($kT_z$)
for an electron temperature of 1.5~keV (Masai 1994).
The horizontal dashed lines represent 90\% errors of the observed values.
  \label{fig:ratio}}
\end{figure}

\acknowledgments
The authors thank the anonymous referee for the valuable comments on our previous draft.
We are grateful to Takeshi Tsuru, Hironori Matsumoto, Hideki Uchiyama, Asami Hayato, 
and the W49B team members for their constructive suggestions.
We especially thank Masaomi Tanaka and Nozomu Tominaga for the dedicated support 
in giving us very helpful advice and improving the draft. 
M.Ozawa is a Research Fellow of Japan Society for Promotion of Science (JSPS). 
This work is partially supported by the Grant-in-Aid for the Global COE Program 
"The Next Generation of Physics, Spun from Universality and Emergence", 
Challenging Exploratory Research (KK), 
and Young Scientists (HY) 
from the Ministry of Education, Culture, Sports, Science and 
Technology (MEXT) of Japan.

\clearpage
\end{document}